\documentclass{ieeeaccess}
\usepackage{cite}
\usepackage{amsmath,amssymb,amsfonts}
\usepackage{algorithmic}
\usepackage{graphicx}
\usepackage{textcomp}

\usepackage{bm}
\makeatletter
\AtBeginDocument{\DeclareMathVersion{bold}
\SetSymbolFont{operators}{bold}{T1}{times}{b}{n}
\SetSymbolFont{NewLetters}{bold}{T1}{times}{b}{it}
\SetMathAlphabet{\mathrm}{bold}{T1}{times}{b}{n}
\SetMathAlphabet{\mathit}{bold}{T1}{times}{b}{it}
\SetMathAlphabet{\mathbf}{bold}{T1}{times}{b}{n}
\SetMathAlphabet{\mathtt}{bold}{OT1}{pcr}{b}{n}
\SetSymbolFont{symbols}{bold}{OMS}{cmsy}{b}{n}
\renewcommand\boldmath{\@nomath\boldmath\mathversion{bold}}}
\makeatother

\def\BibTeX{{\rm B\kern-.05em{\sc i\kern-.025em b}\kern-.08em
    T\kern-.1667em\lower.7ex\hbox{E}\kern-.125emX}}

\def\sir{\langle I\rangle}

\begin{document}
\history{Date of publication xxxx 00, 0000, date of current version xxxx 00, 0000.}
\doi{0}

\title{Targeted Quarantine Strategies Mitigate SIS-model Cyber Epidemics on Complex Networks}
\author{\uppercase{Ho Yuen Wong}\authorrefmark{1}, AND
\uppercase{Tak Shing Tai}\authorrefmark{1}}

\address[1]{Department of Electronic Engineering and Computer Science, Hong Kong Metropolitan University, Homantin, Hong Kong SAR, China}

\markboth
{Wong \& Tai: Preparation of Papers}
{Wong \& Tai: Preparation of Papers}

\corresp{Corresponding author: Tak Shing Tai (e-mail: stai@hkmu.edu.hk).}

\begin{abstract}
The rapid expansion of highly connected and heterogeneous digital infrastructures has fundamentally altered the dynamics of cyber epidemics, making it increasingly important to understand how malware propagates under realistic structural and policy conditions. In this study, we employ an SIS epidemic model on complex networks to investigate virus spreading under a range of cybersecurity scenarios, including minimally protected environments, anti-virus, vulnerable legacy systems, mixed security systems, rapidly evolving virus, and targeted quarantine of key nodes. Using simulations on scale-free topologies, we characterize how infection prevalence evolves over time and identify conditions under which outbreaks become explosive versus controlled. Our results show that vulnerable and rapidly evolving virus scenarios lead to fast and almost complete compromise, while antivirus and mixed-security configurations significantly reduce peak infection levels. Most importantly, we demonstrate that targeted quarantine of a small fraction of structurally important nodes can both delay the onset of large scale outbreaks and lower steady state infection, revealing a threshold in the fraction of quarantined nodes which required to alter spreading of computer virus. These findings highlight the critical role of network topology and selective protection in enhancing the resilience of cyber-physical infrastructures.
\end{abstract}


\titlepgskip=-21pt

\maketitle

\section{Introduction}\label{sec1}

\IEEEPARstart{T}{he} rapid expansion of highly connected digital ecosystems has fundamentally changed the landscape of cyber threats. Modern networks now interlink traditional IT infrastructures with cloud platforms, mobile devices, industrial control systems, and billions of IoT sensors and actuators. This ubiquitous connectivity increases the attack surface dramatically, providing adversaries with more entry points and pathways for lateral movement. Malware campaigns can now spread at unprecedented speed, often propagating automatically without requiring direct human interaction. Moreover, critical infrastructures such as energy grids and transportation systems are increasingly dependent on interconnected and often heterogeneous systems, where a single compromised component can trigger cascading failures \cite{JiangJeusfeldSandahl2023}. The cyber attacks can also cause physical damage and large scale service disruptions. At the same time, the complexity and opacity of these networks make it difficult to maintain complete visibility to monitor all communication channels and isolate compromised segments. As a result, understanding how cyber threats emerge and propagate in such highly connected environments has become a challenge for cybersecurity research and a key motivation for simulation of virus spreading in complex networks.

Moreover, some organizations rarely have a single, well-defined perimeter. Cloud services and any third-party integrations fragment the boundary into a collection of shifting, partially controlled edges, which makes it difficult to enforce consistent policies and to maintain a clear notion of inside versus outside. As a result, traditional models struggle to predict or contain the spread of malware in realistic environments, where trust assumptions can be violated and connectivity patterns are highly heterogeneous. These limitations motivate a shift towards models and tools that explicitly capture the complexity of real networks and the dynamics of malware propagation. Simulations in complex networks provide a way to study how viruses spread under different structural and policy conditions. By representing systems as graphs and simulating infection processes over time, researchers can examine how factors such as degree distribution, network structure and topologies, and key nodes influence propagation speed, infection reach, and outbreak thresholds  \cite{CrockerStrombom2023,ZhouLiZhang2025,Zapperietal2025}. Also, the simulations can make it possible to conduct controlled experiments that would be infeasible and dangerous on production systems. For example, the simulations can compare how fast a given malware strain spreads in a scale-free network versus a more uniformly connected topology. Furthermore, researchers can evaluate different defense strategies to measure their effect on infection. Another advantage of simulation is its ability to incorporate heterogeneity and dynamics, since real networks contain nodes with different security postures and patch levels. Network connectivity may also change over time as devices connect and disconnect, As a result, simulations can then model these aspects explicitly, so they deepen theoretical understanding of how cyber epidemics behave in complex environments, revealing critical thresholds and structural vulnerabilities. On the other hand, they provide practical guidance for security architecture design, helping to identify which controls such as  stronger authentication and access control, segmentation of high-degree hubs can effectively disrupt malware spread. 

In this study, we reveal how the topology and interdependence patterns of critical infrastructure networks fundamentally shape their cybersecurity resilience. By modeling computer systems and their communication links as a complex network, we quantify how attacks on a small subset of strategically important nodes can propagate to trigger cascading failures across multiple subsystems. We adopt an SIS model on scale-free networks to investigate how different security configurations and intervention strategies shape cyber-epidemic dynamics. In particular, we focus on the role of targeted quarantine, where only a small fraction of structurally central nodes can be isolated due to resource constraints. Unlike immunization or uniform hardening, targeted quarantine exploits network topology considering the connectivity of individual nodes to disrupt critical transmission of computer virus. We compare this strategy against different scenarios, including minimally protected environments, antivirus-enhanced networks, legacy-dominated systems, mixed security postures, and settings with rapidly evolving malware. Our results shed light on the number of connections in the global structure influences the overall vulnerability of the infrastructure. 

By analyzing both node-based infection prevalence and a degree-weighted infection metric over time, we quantify how these scenarios differ in terms of outbreak speed, peak infection level, and persistence. This allows us to identify practical thresholds in the fraction of quarantined nodes required to meaningfully alter epidemic trajectories, and to clarify when topology-aware quarantine can outperform different defense strategies.

\section{Literature review}\label{sec:literature_review}
Research on epidemic spreading in complex networks has developed rapidly over the past two decades, providing a foundational framework for understanding both biological and cyber contagion processes. On the other hand, most existing studies related to cybersecurity assume uniform protection which focus on static immunization of nodes or consider coarse-grained isolation of entire sub-networks so they provide limited insight into the role of selective and quarantine at the node level. Thus, the epidemic models can help improve the understanding of the spreading of virus related to cybersecurity.

\subsection{Epidemic Models for Malware Propagation}
Epidemic models originally developed in mathematical epidemiology has been widely adapted to describe malware propagation in computer and communication networks \cite{Bailey1975}. In their simplest form, these models classify each node into a small number of states and define transition rules between them \cite{Cohen1987,IlgunKemmererPorras1995,LiSaad2024}. The Susceptible–Infected (SI) model captures scenarios where once a device is compromised, it remains infected indefinitely, approximating fast-spreading worms in the early stages of an outbreak or systems without effective remediation. The Susceptible–Infected–Susceptible (SIS) models extend this by allowing infected nodes to be clean but vulnerable again, reflecting environments where patches are incomplete or users repeatedly disable protection \cite{VanMieghem2011,NakamuraMartinez2019,BalzottiDOvidioLaiLoreti2021}. The Susceptible–Infected–Removed (SIR) models include a removed state representing nodes that are patched or a recovery state that is immune to reinfection \cite{Newman2002}. Furthermore, Susceptible–Exposed-Infected–Removed (SEIR) models insert an exposed state to represent latent periods where malware has been delivered but not yet activated, which is relevant for ransomware with countdown timer, or multi-stage intrusion campaigns \cite{TomeOliveira2011,Viguerieetal2021,DzamicMarkovic2026}.

In the cyber context, the parameters of these models such as infection rate, recovery rate, and immunity duration are mapped to technical and organizational factors. The scanning speed, exploiting success probability, patch deployment, incident response, and user behavior are some factors that can be considered in the models. Analytical treatments of SIS or SIR models on networks often focus on deriving epidemic thresholds such as the critical values of the infection rate. These thresholds can depend on network structure and heterogeneity in the complex networks \cite{FengDingHuangZhang2016,KrauseKurowskiYawarVanGorder2018,LiSaad2021}. Some models incorporate heterogeneous infection probabilities for different node types such as servers and clients, time-varying rates, or additional states for detection \cite{QuWang2017,Clancy2018,ChenLiuYuLi2021}. Thus, these models provide some insights to the virus spreading in cyber networks. This gap motivates a systematic investigation of targeted quarantine policies and their impact on the mitigation of SIS-type cyber epidemics on complex networks.

\subsection{Network topology} 
Different algorithms can be used for generating random networks, including Barabási–Albert (BA), Erdős–Rényi (ER) model and the Watts–Strogatz (WS) model. The BA model can generate some human-made systems such as Internet and World Wide Web \cite{BA1999,XuWang2011,GhavasiehDeDomenico2024}. Some studies analyze the network structure of systems such as the connections between devices since networks or graphs have specific topological properties \cite{Pastor-SatorrasVespignani2001,KorngutAssaf2025}. For instance, Internet-topologies, like server–client structure, exhibit a highly heterogeneous connectivity pattern, in which a small number of nodes act as hubs with a very high degree of connectivity while the vast majority of nodes have only a few connections \cite{ZhouMondragon2004,MotamediRejaieWillinger2015,TaiYeung2021}. Moreover, small world networks combine high local clustering with short average path length, meaning that any two nodes are connected by relatively few hops and they can join together by a small number of links so these structural features have a direct impact on how malware and other cyber threats spread \cite{VentrellaPiroGrieco2018,YangHuangLi2019}. In scale-free networks, highly connected hubs act as super-spreaders \cite{Massadetal2008}. When they are compromised, they can infect many neighbors rapidly, dramatically accelerating propagation \cite{CohenHavlinbenAvraham2003,Chernikovaetal2022}. At the same time, such networks are robust to random node failures but extremely vulnerable to targeted attacks on hubs so it implied that protecting or segmenting high-degree nodes is crucial for containment \cite{ShenNguyenXuanThai2013,Ugurlu2025,WangDengWangKurths2025}. Moreover, small world properties can increase the rate of propagation, allowing malware to traverse the network quickly via a few long-range connections, even if most nodes are only locally connected \cite{Zanette2001,ZekriClerc2001,Tianetal2025}. However, some social network structure can be slow for the spread as infections may remain confined within a community for some time, but once a bridging node or gateway is compromised, the malware can jump to new modules and trigger cascading outbreaks \cite{pareetal2023,Genicot2022,Coyac-TorresSidorovAguirre-AnayaHernandez-Oregon2023}.

Understanding these topological characteristics is therefore important to designing defense strategies and simulation models. Studies of epidemic processes on complex networks show that critical thresholds of infection rates depend strongly on the degree distribution and network heterogeneity \cite{Qiuetal2022,MartinCastellaHamelin2026}. For the models in highly heterogeneous, scale-free networks, the thresholds may even vanish, implying that arbitrarily small infection rates can sustain persistent epidemics \cite{AcharyaMondalUpadhyayDas2024,ZhangDengRuanZhang2026}. For cybersecurity, this means that the assumptions of homogeneous mixing can underestimate risk, so it will misguide defense planning \cite{JingJinPeng2018,AletaFerrazMoreno2020}. By explicitly modeling degree heterogeneity and clustering, the simulations can identify critical nodes and evaluate the impact of segmentation to reveal non-intuitive effects on the resilience of networks.

\subsection{Cybersecurity Propagation Models}
The models related to cybersecurity in complex networks can refine generic epidemic frameworks to capture the spreading behaviors of ransomware in real networks. The models often incorporate explicit scanning strategies such as IP scanning and protocol-level constraints, which strongly influence propagation speed and coverage \cite{xinetal2015,SemerciCemgilSankur2018}. Some models can distinguish node roles in enterprise environments and critical infrastructures \cite{JiangJeusfeldSandahl2023}. Moreover, some defense-oriented models extend these approaches by adding explicit mechanisms for detection, patch propagation and quarantine, which can partially change network configuration \cite{ChenXiaPerc2024,ZhangDengRuanZhang2026,HerttuainenKuikkaKaski2026}. For example, immunization processes can represent automated patch rollout or policy updates, while quarantine states capture host isolation by endpoint detection and response or network access control \cite{DzamicMarkovic2026,LiZhangZhou2025}. Those measures may encode continuous authentication and authorization as reduced effective infection rates on protected edges on allowable infection paths between security zones with least-privilege access as smaller reachable neighborhoods for compromised identities. 

Furthermore, some studies consider and optimize the cost of security to explore how defenders should allocate resources such as hardening hubs and deploying additional verification checks to minimize outbreak size or expected loss \cite{Maietal2021}. Some studies can quantify how architectural choices (such as segmentation granularity or policy strictness) shift epidemic thresholds and slow lateral movement that limit the blast radius of successful compromises in complex networks \cite{DaXuZhao2019}.

\section{Methodology}
We employ a model with $N$ nodes, denoted by $i=1,\dots,N$, each node is connected with $k_i$ neighbors, generated by using the BA model. There are different states of different nodes. Initially, all nodes are susceptible with $S$ state so $S_i(t=0) = 1$ for $\forall i$ and the total number of node in $S$ state at time $t$ is $S(t)$ and it can be obtained by

\begin{align}
    S(t) = \sum_{i=1}^N S_i(t).
\end{align}

\begin{figure}
\centering
\includegraphics[ width=0.6\linewidth] {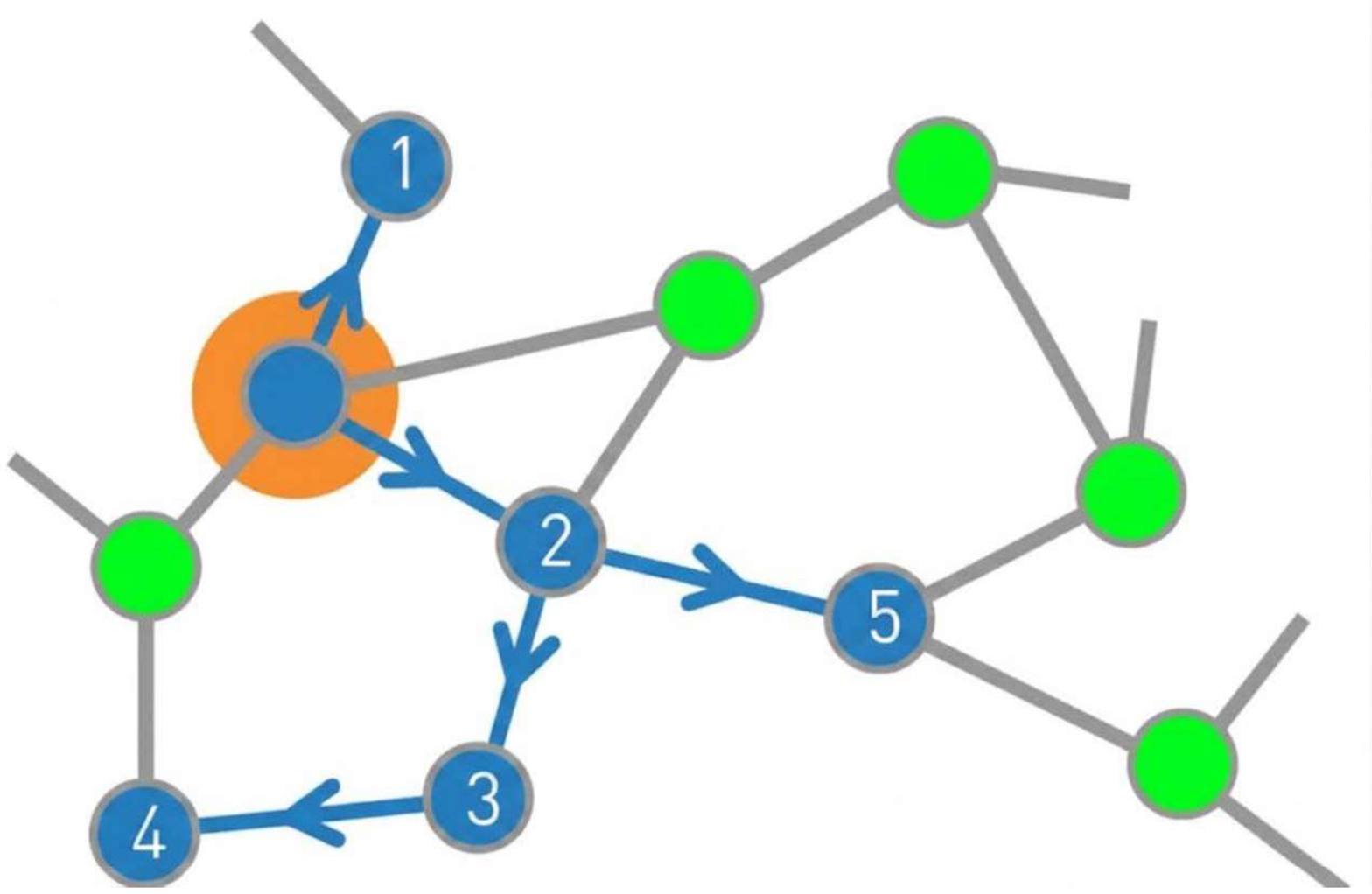}
\caption{
An example of computer virus spreading. The original resource of computer virus, in the orange spot, connected with different nodes with $k=3$. It infected the nearby nodes $i=1$ and $i=2$. The node $i=2$ with $k=4$ further infected node $i=3$ and $i=5$. The node $i=3$ infected the node $i=4$. It showed how the computer virus spread in the network.
}
\label{fig_sis_model}
\end{figure}

When $t=0$, the susceptible nodes are free of virus. However, at the beginning of each simulation, $n$ nodes are selected for containing the source of virus spreading, so $S(0)=N-n$. When the nodes are infected, they become infected nodes from $S \rightarrow I$. An example is shown in figure~\ref{fig_sis_model}. The total number of infected nodes, denoted by $\lambda(t)$, increase when $t$ increase. When $t=0$, $\lambda(0)=n$. When $S$ nodes are connected with or communicating with $I$ nodes, the $S$ nodes can become $I$ nodes, depending on the probability $\beta$, since the $I$ nodes can then transmit virus to their neighboring nodes. When it is infected, more nodes become infective $I$. Furthermore, the $I$ node can recover and become susceptible $S$ again that $I \rightarrow S$, with the probability $\gamma$.  During time $t$, the total infected node $I$ becomes

\begin{align}
    \lambda(t) = \sum_{i=1}^N I_i(t).
\end{align}

$A_{ij}$ describe whether node $i$ can interact with node $j$, for $\forall i,j$, such that

\begin{eqnarray}
\label{eq_node_connect}
A_{ij} =
\begin{cases}
1, &\mbox{if nodes $i$ and $j$ are connected}
\\
0 , &\mbox{otherwise.}
\end{cases}.
\end{eqnarray}

In the networks, $A_{ij} = A_{ji}$. For individual node, we can identify the key nodes by the sum of infection contributions from all infected neighbors $j$ that are connected to node $i$ by obtaining the value of $\sum_j A_{ij} I_j(t)$. We draw a random number $\eta$ to determine whether any node i, where $S_i(t-1)=1$, will be infected at time $t$. On each connection with infected nodes, the $I_i(t)$ at time $t$ become 

\begin{eqnarray}
\label{eq_pro_I}
I_i(t) =
\begin{cases}
1, &\mbox{if $\eta<\beta$}
\\
0 , &\mbox{otherwise.}
\end{cases},  \mbox{where } S_i(t-1)=1, \forall i,t.
\end{eqnarray}
When $I_i(t)=1$, $S_i(t)=0$. Since not all nodes will be connected to all other nodes, the spreading rate may be slower than some traditional SIS/SIR models. Hence, we rewrite the net rate of the change of the total number of I at $t$ 

\begin{align}
\frac{d \lambda(t)}{dt}
&= \sum_{i=1}^N \frac{d I_i(t)}{dt}\\
\label{eq_net_rate_i}
&= \beta \sum_{(ij)}  A_{ij} S_j(t) \sum_{(ij)}\frac{A_{ij} I_j(t)}{N} 
- \gamma \sum_{i=1}^N I_i(t).
\end{align}
$\beta$ and $\gamma$ are distinct values in different scenarios. Further discussion is included in section~\ref{sec:different_scenario}.
When $t \rightarrow \infty$, the state of all nodes become $I$, so we determine the infected rate and spreading rate within time limit, as mentioned in section~\ref{sec:quantities_interest}.

\subsection{Scenarios with quarantine state}
As the infection is mostly from infected neighbors, the protection to neighbor is vital. Hence, to stop the virus spreading, we introduce a state $Q$ that the node $i$ cannot be infected  (i.e. cannot be change from $S \rightarrow I$). The number of nodes that are quarantined, denoted by $n_Q$, is between $0 \le n_Q \le 1$, and 

\begin{align}
Q_i(t) = 1,\quad I_i(t) = 0,\quad S_i(t) = 0 \quad \forall t, i \in Q.
\end{align}

Hence, at time $t$, the total number of nodes $N$ becomes
\begin{align}
N = S(t) + \lambda(t) + Q(t)
\end{align}

Since protecting all nodes from the virus is costly, we assume that only a fraction of nodes, denoted by $p$, can be endowed with enhanced protection with

\begin{align}
    p = \frac{n_Q}{N},
\end{align}
where $p \ll 1$. Only a small subset of nodes is assumed to be protected. These $Q$ nodes are intended to represent critical vertices or servers in the network, for which enhanced defensive measures are deployed to prevent them from acting as major sources for infection. We quantify the fraction of such key nodes that must be quarantined to achieve a reduction in spreading. We reveal how to select the number so as to maximize the mitigation of spreading.

For the scenarios with $Q$ state, we rewrite the Eq.\ref{eq_net_rate_i} to

\begin{align}
\frac{d \lambda(t)}{dt}
= \beta \sum_{(ij)}  A_{ij} S_j(t) \sum_{(ij)}\frac{A_{ij} I_j(t)|Q_i(t)-1|}{N} \nonumber\\
- ~\gamma \sum_{i=1}^N I_i(t)|Q_i(t)-1|
\end{align}

\subsection{Different scenarios for simulation}
\label{sec:different_scenario}

We consider six representative cybersecurity scenarios in our SIS modeling simulation, as shown in table~\ref{tab:different_scenario}. The first case, named default, represents a standard network with only basic protections and insufficient expertise to recover compromised systems. This is captured by moderate infection and recovery rates with $\beta=0.15$ and $\gamma=0.03$. The second case, namely anti-virus, corresponds to a heavily patched and well-maintained network, in which improved defenses reduce the infection rate and accelerate recovery with $\beta=0.03$ and $\gamma=0.08$. In contrast, the third case, namely vnpatched/vulnerable networks, is a model in legacy or unpatched endpoint environments, where the lack of updates and protections leads to a high infection rate and a low recovery rate with $\beta=0.35$ and $\gamma=0.03$. The fourth case, namely targeted quarantine, is a scenario explicitly isolates the most connected nodes based on node connectivity. It introduces a quarantined (Q) state for selected nodes to limit propagation through critical hubs with same $\beta$ and $\gamma$ as in the first case. The fifth case, namely mixed security level, is a case assumes heterogeneous protection levels, with three classes of nodes including hardened, standard, and legacy.  Each class is characterized by distinct $\beta$ and $\gamma$ with a fraction $u$ of nodes hardened, a fraction $v$ of nodes at standard security, and the remaining fraction $w$ operating as legacy systems, where
\begin{align}
    u + v + w = 1,
\end{align}
thereby reflecting non-uniform security postures. 

Finally, the case, namely rapidly evolving virus, is a scenario captures adaptive malware that becomes more effective over time. When the overall infection level exceeds $50\%$, the virus progressively increases its infection rate and decreases the recovery rate at each time step by $\Delta \beta=+0.05$ and $\Delta \gamma=-0.02$, modeling an arms race where the threat evolves in response to widespread compromise.

\begin{table*}
    \centering
    \begin{tabular}{|l|l|l|}
    \hline
    \textbf{Case} & \textbf{Description} & \textbf{Features} \\ \hline
    \begin{tabular}[c]{@{}l@{}}Default \end{tabular} & \begin{tabular}[c]{@{}l@{}}Standard connection with insufficient skills  \\ and minimal cybersecurity supports\\ to recover the systems\end{tabular} & Fair $\beta$ and $\gamma$ \\ \hline
    \begin{tabular}[c]{@{}l@{}}Anti-virus\end{tabular} & \begin{tabular}[c]{@{}l@{}}A heavily patched \\ and well-maintained network\end{tabular} & Low $\beta$ and high $\gamma$ \\ \hline
    \begin{tabular}[c]{@{}l@{}}Unpatched\end{tabular} & \begin{tabular}[c]{@{}l@{}} Unpatched or vulnerable systems \\with legacy or unpatched endpoint environment \end{tabular} & High $\beta$ and low $\gamma$ \\ \hline
    \begin{tabular}[c]{@{}l@{}}Targeted\\ Quarantine\end{tabular} & \begin{tabular}[c]{@{}l@{}}Using degree centrality \\ to isolate the most connected nodes\end{tabular} & Q state in the nodes \\ \hline
    \begin{tabular}[c]{@{}l@{}}Mixed \end{tabular}& \begin{tabular}[c]{@{}l@{}}Mixed security levels include \\three different security levels for nodes: \\ $u$ hardened, $v$ standard and $w$ legacy\end{tabular} & each with unique $\beta$ and $\gamma$ \\ \hline
    \begin{tabular}[c]{@{}l@{}}Rapidly \\Evolving\end{tabular} & \begin{tabular}[c]{@{}l@{}}Rapidly evolving virus is \\ virus that will become smarter overtime \end{tabular}& \begin{tabular}[c]{@{}l@{}}When the infection rate is over $50\%$, \\ $\beta$ increase and $\gamma$ decrease in each time $t$\end{tabular} \\ \hline
    \end{tabular}
    \caption{Different scenarios in the simulations. Six cases with different $\beta$ and $\gamma$ show different cybersecurity settings}
    \label{tab:different_scenario}
\end{table*}

\subsection{Quantities of interest}
\label{sec:quantities_interest}
The simulations start at $t=0$. When $t=T$, we obtain the state of all nodes and calculate the summarized infected rate, denoted by $\sir$, as
\begin{align}
    \sir = \frac{1}{N}\sum_{i} |I_i(T)|.
\end{align}

The infected rate at time $t$, denoted by $I(t)$ measures the fraction of the number of infected nodes in the network. 
\begin{align}
    I(t) &= \frac{1}{N}\sum_{i} |I_i(t)|\nonumber\\
         &= \frac{1}{N}\lambda(t).
\end{align}
For each node $i$, $|I_i(t)|$ denotes whether the node is infected at time $t$. 

The weighted infected rate at time $t$, denoted by $\omega(t)$ measures the fraction of edge-weighted infection in the network. 
\begin{align}
    \omega(t) = \frac{1}{N\sum_i k_i }\sum_{i} |I_i(t)|\times k_i.
\end{align}
$k_i$ is its degree and $|I_i(t)| \times k_i$ counts the number of incident edges attached to infected nodes. It emphasizes infections at highly connected nodes, capturing how many nodes are infected and how structurally important those infections are for potential spread.


\section{Results}\label{sec:result}
In this section, we will show the results of different quantities of interest, including $\sir$, $I(t)$ and $\omega(t)$ between different cases with distinct $\beta$ and $\gamma$.

\subsection{Impact on different scenarios for infection spreading over time}
\begin{figure*}
\centering
\includegraphics[ width=0.9\linewidth]{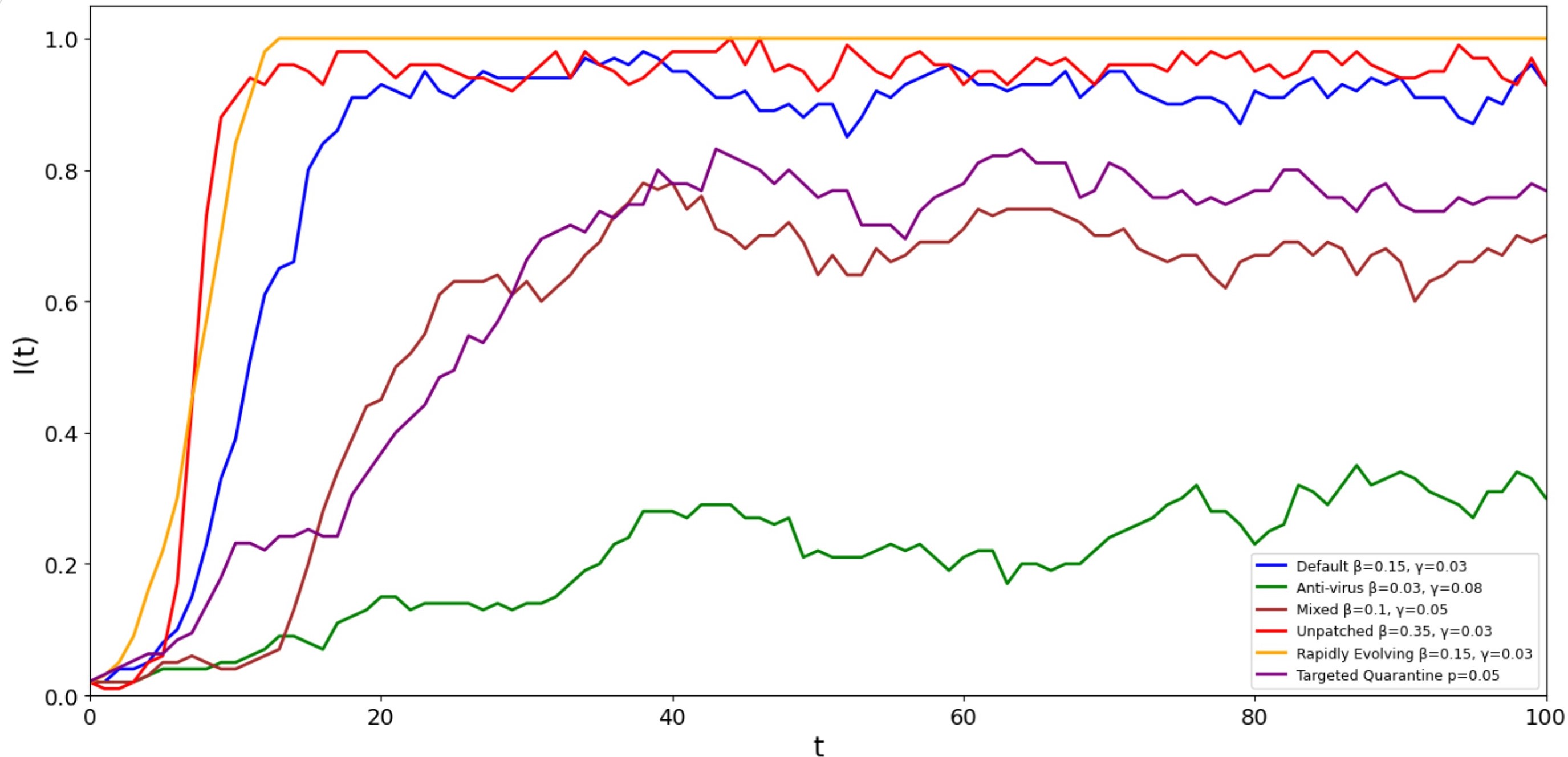}
\caption{
The simulation results of $I(t)$ as a function of $t$. Results for various cases stated in table~\ref{tab:different_scenario} are obtained. We can see that both $I(t)$ increases with $t$. For the cases with default setting, vulnerable systems and rapidly evolving virus, the infection is serious. For the cases with mixed security and targeted quarantine, the infection is less serious. For the cases with anti-virus, the infection is much mild. We can see that the mixed security, targeted quarantine and anti-virus can be used to preventing the computer virus from disrupting the whole network.
}
\label{fig_diffent_scenario}
\end{figure*}

We first examine how different scenarios in section~\ref{sec:different_scenario} impact the infection spreading rate. As shown in figure~\ref{fig_diffent_scenario}, the infections generally start from near zero at $t=0$ and rise over time, and the rises rapidly to a high steady level near 100\% of nodes infected in some cases where $\sir=1$. For the cases with the default setting, The infected nodes increase gradually but steadily from $t=0$ to $t=10$. The infection reaches $45\%$ of nodes at $t=10$. The infection rate keep increasing from $t=10$ to $t=30$ and around $75\%$ nodes are infected. The infection stabilizes into a high plateau and hovers around $95\%$ infection for the remainder of the simulation (i.e. $\sir=0.95$), infecting most of the network most of the time. it shows that this setting permits widespread of virus, if there are no additional measures are conducted. For the cases with anti-virus, the infection rate rises much more slowly from zero than the others. At $t=10$, only $7\%$ of nodes are infected. From $t=10$ to $t=40$, the infection rate keep increasing but the slope is low. After $t=40$, the infection seems to fluctuates between $20\%$ to $30\%$, showing that the strong anti-virus controls can minimize infection prevalence. For the cases with unpatched/vulnerable systems, the infection rate is the steepest, At $t=5$, near half of the nodes are infected and at $t=10$, around $90\%$ of nodes are infected. The cases suffer fast and nearly complete compromise. Similarly, for the cases with evolving virus, the number of infected nodes also increase very fast. At $t=4$, around $90\%$ of nodes are infected. This is the worst-case scenario where the virus rapidly conquers the network. A rapidly evolving virus is able to overcome defenses to produces an explosive outbreak.

\subsubsection{Comparison between vulnerable networks and antivirus networks}
\begin{figure*}
\centering
\includegraphics[ width=0.8\linewidth]{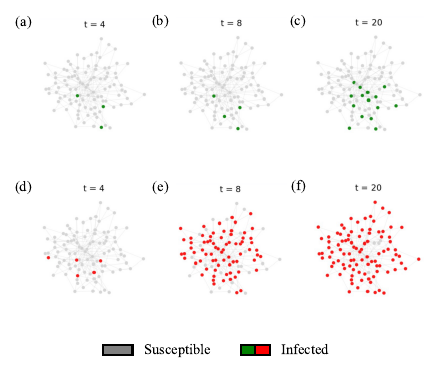}
\caption{
An example of snapshots susceptible and infected nodes with the cases with anti-virus in (a)-(c), and the cases with vulnerable networks in (d)-(f). The susceptible nodes are in grey color, whereas the infected nodes are in green or red color, representing the infected nodes in the case with anti-virus  and the case with anti-virus respectively. At $t=4$, the number of infected nodes in (a) and (d) are similar. However, at $t=8$, the number of infected nodes in the case with anti-virus in (b) is much smaller than the the number of vulnerable networks in the case with anti-virus in (e). At $t=20$, the number of infected nodes in the case with anti-virus in (c) increase a lot, but it is smaller than the case with vulnerable networks.
}
\label{fig_vul_anti}
\end{figure*}

As shown in figures~\ref{fig_vul_anti}(a) and \ref{fig_vul_anti}(b), when $t=4$, the number and distribution of infected nodes in the cases with antivirus are still comparable to those in the cases with vulnerable networks. At this early stage, the virus has not yet had sufficient time to fully exploit structural differences, so both scenarios exhibit a similar initial growth of infections.
However, when $t=8$, a clear divergence emerges. In the cases with vulnerable networks, the infection spreads rapidly and a large fraction of nodes become compromised, indicating an early onset of a widespread outbreak. In contrast, in the cases with antivirus, the spreading is effectively suppressed, and only a very small fraction of nodes are infected. This suggests that even a moderate level of protective capability can significantly delay the contagion process and reduce the infection level.
When $t=20$, the contrast becomes more pronounced. In the cases with vulnerable networks, almost all nodes are infected, implying a near-complete breakdown of security that illustrating the unmitigated malware propagation in critical infrastructures. For the cases with antivirus, the virus still begin to spread more noticeably, but the growth is much slower and the overall infection level remains substantially lower than in the cases with vulnerable networks. This gradual increase reflects the competing dynamics between infection and recovery processes introduced by the antivirus mechanisms. Moreover, it highlights the crucial role of defensive measures in reshaping the temporal evolution of virus spreading of cybersecurity problem.

\subsubsection{Challenge on mixed security}
For the cases with mixed security, The infection rate grows moderately fast. It is faster than the cases with anti-virus but much slower than the cases with vulnerable networks and evolving virus. At $t=20$, around $50\%$ of nodes are infected. After $t=30$, it approaches a plateau at around $70\%$ of nodes infected. It shows that the cases with mixed security yields an intermediate outcome. When a fraction of the networks is infected, some nodes can still prevent themselves from infection. It demonstrates that partial hardening helps but still leaves a large infection reservoir. In this case, we set $u=0.3$ and $v=0.4$ for the security levels for nodes. It shows that the heterogeneity in security improves the situation relative to uniformly weak defenses. The improvement takes part of the network’s security that it can avoid the near-total compromise.

\subsection{Importance on quarantine nodes}
For the cases with targeted quarantine, we set $p=0.05$ and it shows a slower rise in the number of infected nodes over time. At $t=10$, around $40\%$ of nodes are infected, which is lower than the cases with the default setting. At $t=20$, the infection rate is around $70\%$, which is also lower than the cases with the default setting. From $t=30$, the infection rate keep fluctuating between $60\%$ to $70\%$ and it never reaches the maximum. We found that adding targeted quarantine to the systems can reduce the maximum infection level.

Similar to the cases with mixed security, the infection rate in the network in the cases with targeted quarantine reduced. We can see that even though $p$ is small, the outcome of preventing infection can be similar to the cases with mixed security. The effort of mixed security can be much higher to prevent the infection as $u=0.3$. However, identifying the key nodes with higher connectivity is vital for reducing the infection by the cases with targeted quarantine.

\begin{figure*}
\centering
\includegraphics[ width=1.0\linewidth] {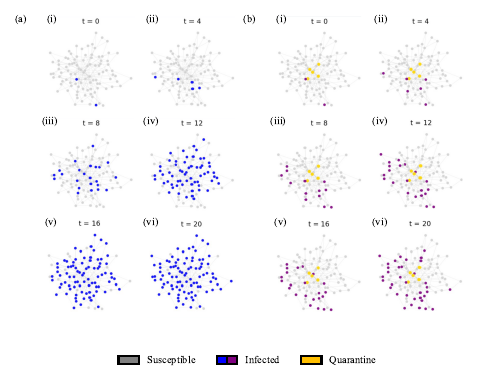}
\caption{
An example of snapshots susceptible, infected and quarantine nodes with the cases with default setting in (a), and the cases with targeted quarantine $p=0.05$ in (b). The susceptible nodes are in grey color, the infected nodes are in blue or purple color, representing the infected nodes in the case with default setting and the case with targeted quarantine respectively, and the quarantine nodes are in yellow color. The sequence is from (i) to (vi). We can see that at $t=20$, the number of infected nodes in the case with default setting in (a)(vi) is much larger than that of the case with targeted quarantine in (b)(vi).
}
\label{fig_targetQ}
\end{figure*}

In figure~\ref{fig_targetQ}, we compare the case with the default setting with the case with targeted quarantine and $p=0.05$. At $t=0$, 5 nodes are selected to become quarantine nodes in the case with targeted quarantine. At $t=4$, as shown in figures~\ref{fig_targetQ}(a)(ii) and figures~\ref{fig_targetQ}(b)(ii), the number of infected nodes are similar. However, when $t$ increases, the difference become observable. When $t=8$, the number of infected nodes in the case with targeted quarantine is slightly smaller than that in the case with the default setting. When $t=12$, it is clearly shown in figures~\ref{fig_targetQ}(a)(iv) and figures~\ref{fig_targetQ}(b)(iv) that the number of infected nodes is smaller in the case with targeted quarantine. The number of infected nodes keeps smaller when $t=16$ and $t=20$, from the figures~\ref{fig_targetQ}(a)(v) and figures~\ref{fig_targetQ}(b)(v), and figures~\ref{fig_targetQ}(a)(vi) and figures~\ref{fig_targetQ}(b)(vi) respectively. We can see that the case with targeted quarantine can minimize the number of infected nodes.

\begin{figure*}
\centering
\includegraphics[ width=0.9\linewidth] {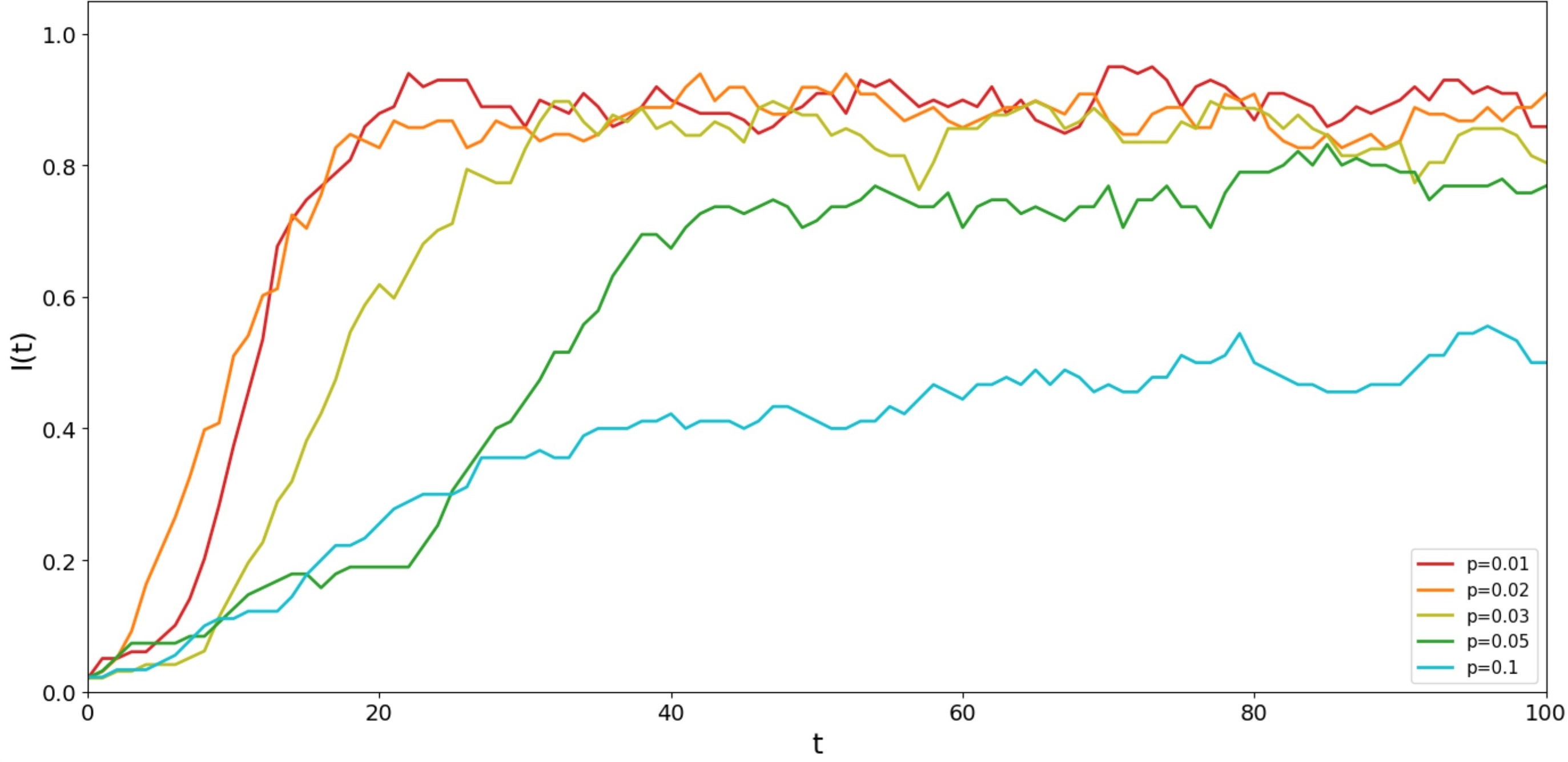}
\caption{
The simulation results of the difference in $I(t)$ between different cases with targeted quarantine, as a function of $t$. Results for various values of $=0.01, 0.02, 0.03, 0.05$ and $0.1$ are obtained. We can see that if we protect more critical infrastructure from the virus spreading which disrupt the major transmission routes, the critical threshold of emerging outbreaks can be delayed.
}
\label{fig_diffQ}
\end{figure*}

Moreover,We study the effect on the quarantine nodes in the network. When $p$ is small, the effect on reducing infected nodes is almost unobservable. We can see that, from figure~\ref{fig_diffQ}, when $p=0.01$ and $p=0.02$, the number of infected nodes increases with $t$ as usual, showing that the protection is far from enough. However, when $p=0.03$, we can see that the critical point for greatly increasing the number of infected nodes shift to larger $t$. It means that the targeted quarantine can delay the $t$ for the critical point. When $p$ increases, the critical points shift to larger $t$. We can see that the targeted quarantine can help delay the occurrence of large group of infection and reduce the number of infected nodes.

We further examine the impact of the fraction of quarantined nodes $p$. When $p$ is small, the effect on reducing infections is almost negligible. As shown in figure~\ref{fig_diffQ}, for $p = 0.01$ and $p = 0.02$, the number of infected nodes increases with time $t$, which is similar to the case with the default setting. It indicates that a low level of quarantine is far from sufficient to alter the overall spreading pattern as the quarantine nodes are too limited to disrupt the main transmission pathways in the network. However, when $p = 0.03$, the onset of the critical point at which the number of infected nodes begins to rise sharply is shifted to larger $t$. This indicates that targeted quarantine can delay the epidemic escalation even when only a small additional fraction of key nodes is protected. As $p$ increases further, this critical point moves to later $t$, and both the infection rate and the peak prevalence of infection are reduced. In other words, it postpones the emergence of large scale outbreaks and lowers the eventual infection level. These observations suggest the existence of a threshold, in terms of $p$. When $p$ is higher, the quarantine measures have more impact on the reduction on the number of infected nodes. This implies that carefully identifying and isolating a relatively small but critical subset of nodes can provide benefits to enhance the resilience of computer virus without requiring universal protection of the entire network.

\begin{figure*}
\centering
\includegraphics[ width=1.0\linewidth]{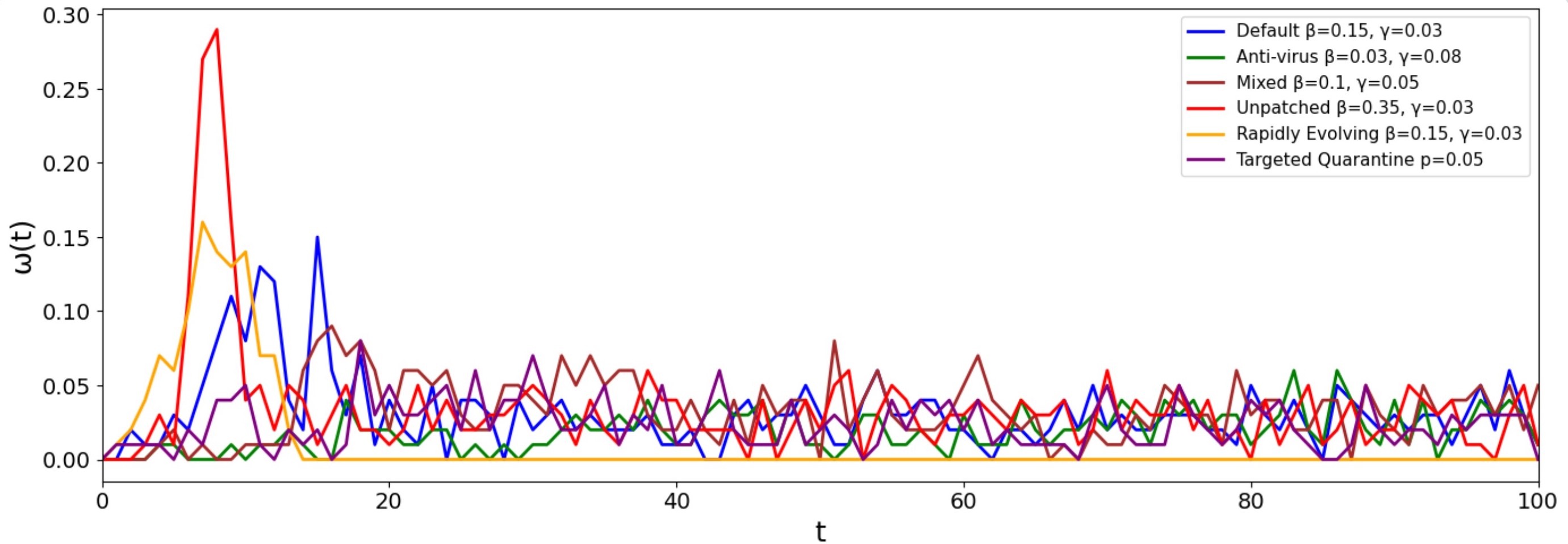}
\caption{
The simulation results of $\omega(t)$ as a function of $t$. Results for various cases stated in table~\ref{tab:different_scenario} are obtained. We can see that there was a large peak of infection in the cases with vulnerable systems and rapidly evolving virus. The peak was delayed in the cases with default setting, but the peak was still observed in the first $10-20$ time steps. For the cases with mixed security, targeted quarantine and anti-virus, we can see that the $\omega(t)$ is low, representing that there are no sudden outbreaks.
}
\label{fig_diffent_scenario_wi}
\end{figure*}

\section{Conclusion}\label{sec13}
This work demonstrates that the interplay between network structure and defensive heterogeneity, including targeted intervention, fundamentally shapes the progression of computer virus spreading in complex systems. By comparing multiple security scenarios within SIS models, we show that minimally protected and legacy-dominated environments are highly susceptible to rapid compromise, especially when confronted with evolving virus. In contrast, antivirus deployment and mixed security systems substantially mitigate infection prevalence. The result that quarantining a relatively small informed subset of nodes can delay epidemic escalation and reduce peak infection is striking as it can reveal a threshold in the quarantined fraction that separates ineffective from impactful intervention. These insights provide quantitative measures for prioritizing protection efforts toward critical infrastructures and nodes, rather than attempting costly and homogeneous hardening of all nodes. Beyond improving understanding of cyber-epidemic behavior on complex networks, our findings offer practical guidance for designing security architectures that leverage topology-aware quarantine from the segmentation of high-degree hubs. The differentiated protection levels can effectively disrupt malware propagation in real-world critical infrastructures.

\section*{Acknowledgment}
The work described in this paper was supported by Hong Kong Metropolitan University Research Grant (Project Reference No. RD/2025/1.27)


\EOD

\end{document}